\documentclass[oneside]{igs} 
\usepackage{cite}

  \usepackage{igsnatbib}

  \ifx\pdftexversion\undefined
    \usepackage[dvips]{graphicx}
  \else
    \usepackage[pdftex]{graphicx}
    \usepackage{epstopdf}
  \fi

\begin{document}

\title[Modelling ice birefringence for neutrino detection]{Modelling ice birefringence and oblique radio wave propagation for neutrino detection at the South Pole}

\author[Jordan]{T. M. JORDAN$^{1,2}$, D. Z. BESSON$^{3,5}$ I. KRAVCHENKO$^{4}$, U. LATIF$^{3}$, B. MADISON$^{3}$, A. NOKIKOV$^{3,5}$, A. SHULTZ$^{3}$}
\affiliation{%
$^1$ School of Geographical Sciences, University of Bristol, UK. \\
$^2$ Department of Geophysics, Stanford University, USA. \\
$^3$ Department of Physics, Kansas University, USA. \\
$^4$ Department of Physics, University of Nebraska, Lincoln, USA. \\
$^5$ NRNU MEPhI, Moscow, Russia. \\

 Correspondence: Tom Jordan and Dave Besson
 $<$tom.jordan@bris.ac.uk$>$, $<$zedlam@ku.edu$>$}

\abstract{The Askaryan Radio Array (ARA) experiment at the South Pole is designed to detect high-energy neutrinos which, via in-ice interactions, produce coherent radiation at frequencies up to 1000 MHz. 
In Dec. 2018, a custom high-amplitude radio-frequency transmitter was lowered into the 1700 m SPICE ice core to provide test sources for ARA receiver stations sensitive to vertical and horizontal polarizations. For these tests, signal geometries correspond to obliquely propagating radio waves from below. The ARA collaboration has recently measured the polarization-dependent time delay variation, and report more significant time delays for trajectories perpendicular to ice flow. Here we use fabric data from the SPICE ice core to construct a bounding model for the ice birefringence and the polarization time delays across ARA. The data-model comparison is consistent with the vertical girdle fabric at the South Pole having the prevailing horizontal crystallographic axis oriented near-perpendicular to ice flow. This study presents the possibility that ice birefringence can be used to constrain the range to a neutrino interaction, and hence aid in neutrino energy reconstruction, for in-ice experiments such as ARA.}
\maketitle
\section{Introduction}
Neutrinos are elementary particles whose interactions with polar ice molecules can be detected from the emission of Cherenkov radiation. The Askaryan Radio Array (ARA) is a neutrino experiment at the South Pole that aims to detect high-energy neutrinos which produce coherent Cherenkov radiation in a radio-frequency band $\sim$ 150-800 MHz \citep{allison2012design}. The ARA detectors are $\sim$ 150 m below the ice surface with the targeted neutrino interactions occurring at ice depths $\sim$ 100-2000 m. The emitted radio waves from the ice-neutrino interaction propagate at oblique angles relative to the vertical direction with the trajectories from the deeper interaction source to the shallower detector often close to horizontal.

Characterization of the radio-frequency (bulk) permittivity of polar ice is critical to optimizing the ARA detector sensitivity and to reconstruct information about ice-neutrino interactions \citep{allison2019measurement,kravchenko2011radio}. The real component of the permittivity relates to the volume of ice visible to the radio receiver array, whilst the imaginary component relates to the attenuation of the radio signal. Due to the presence of ice fabric - the orientation distribution of ice crystals - polar ice behaves as a birefringent medium whereby radio wave polarizations experience different permittivities and propagate at different phase velocities \citep{Hargreaves1978}. Characterization of ice birefringence is important for the ARA experiment as it would facilitate a range estimate from the interaction source to the detector based on polarization time delays at the detectors, and, subsequently aid in neutrino energy reconstruction. Measured polarization time delays using radio test sources have demonstrated that a significant birefringence ($\sim$ 0.1-0.3 $\%$ of the mean refractive index) is present for oblique angle trajectories \citep{allison2019measurement,kravchenko2011radio}. However, a first-principles propagation model that predicts the polarization time delays is yet to be developed.

Since the commencement of the ARA experiment, the SPICE (South Pole Ice Core Experiment) team have collected ice fabric data that extends down to a depth $\sim$ 1750 m \citep{Voigt2017}. Ice fabric is of interest in glaciology as it contains information about ice-sheet flow history \citep{Alley1988} and changes in ice fabric are often correlated with climatic transitions \citep{Kennedy2013}. Polarimetric radar sounding is often used to complement direct sampling of ice fabric \citep{Fujita2006,Li2018,Jordan2019,Drews2012,Dall2010,Brisbourne2019}, and it can be used to reference azimuthal fabric orientation which is generally not recorded directly during ice coring \citep{Wang2002,Faria2010}. Terrestrial radar-sounding systems operate in a similar frequency range  ($\sim$ 50-400 MHz) to the coherent radio-frequency emissions induced by neutrinos. Consequently, effective medium models of ice birefringence and radio propagation developed by the radar-sounding community \citep{Fujita2006,Matsuoka2009} are of direct relevance to understanding the polarization time delays at the ARA experiment.

The primary focus of this study is to use SPICE fabric data to place bounds upon radio-frequency ice birefringence and polarization time delays for oblique radio propagation at the ARA. In Section 2 we motivate the investigation by summarizing oblique radio test-source measurements for polarization time delays, including new measurements from Dec. 2018. In Section 3 we use SPICE fabric data \citep{Voigt2017} and an effective medium framework \citep{Fujita2006} to model the radio-frequency ice birefringence at the South Pole. In Section 4 we construct an oblique radio wave propagation model. In Section 5 we place model bounds upon polarization time delays and compare with the measurements. In Section 6 we discuss the consequences for neutrino source reconstruction and the understanding of ice fabric in the South Pole region.
\section{Radio test-source measurements}
%
\subsection{Overview of the ARA experiment}
The ARA experiment is designed to operate in the 150-800 MHz radio-frequency band, with the goal of detecting impulsive emissions from collisions
of high-energy, extraterrestrial neutrinos with ice molecules. As part of the initial ARA receiver calibration, three `deep' radio-frequency pulsers were deployed at the depths typical of neutrino interactions (two at 1400 m and one at 2450 m depths). More recently, in Dec. 2018 a custom high-amplitude radio-frequency transmitter was lowered into the 1700 m SPICE ice core to provide test sources for ARA receiver stations which are located 1-5 km from the test sources.

In this study we focus upon measurements from ARA receiver stations 2 and 4 (``A2'' and ``A4'', respectively). The radio trajectories to A2 are approximately flow-perpendicular (azimuthal angles to flow $\sim$ 77 $^\circ$ and $\sim$ 79 $^\circ$ respectively for the SPICE and deep pulser sources). The radio trajectories to A4 are approximately flow-parallel (azimuthal angles to flow $\sim$ 1 $^\circ$ and $\sim$ 18 $^\circ$ respectively for the SPICE and deep pulser sources). A plan-view of the experiment in shown in Fig. 1.
\begin{figure}
\centerline{\includegraphics[width=80 mm]{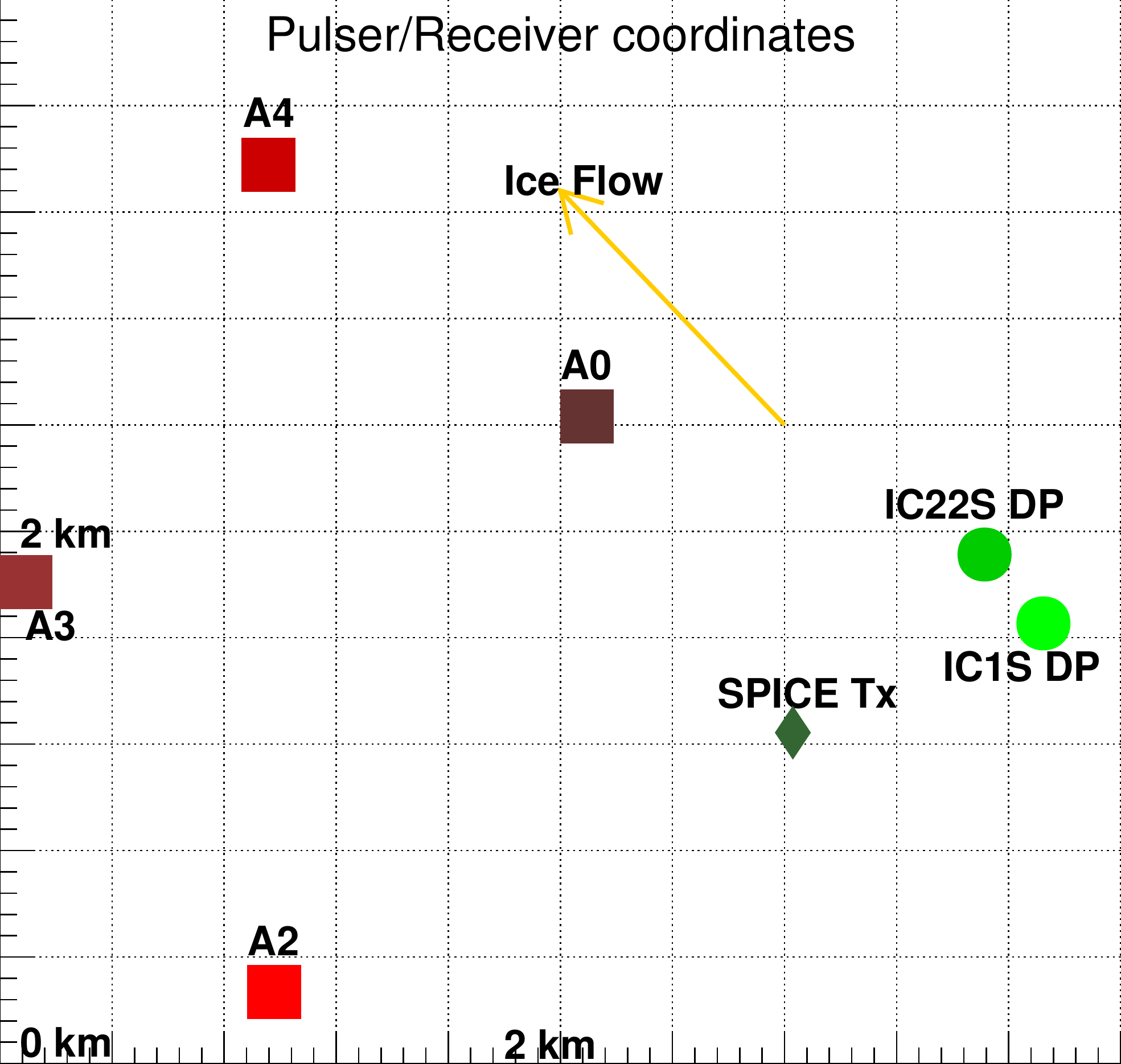}}
\caption{Plan-view of ARA experiment at the South Pole, indicating the location of the receiver stations A2 and A4, the SPICE ice core pulser and the deep pulser sources IC1S DP and IC22S DP, at
a depth of approximately 1400 m. The approximate ice flow direction is also indicated.} 
\label{ARA}
\end{figure}

\subsection{Polarization time delay measurements}
The ARA collaboration has reported time delays for data taken using the
deep pulser sources IC1S DP and IC22S DP \citep{allison2019measurement,tom2019ICRC,allison2019long},
as well as for the SPICE icehole pulser. Most interestingly, they
find signal arrival time differences between horizontally polarized (h) versus vertically polarized (v) signals
arriving from the A2 and A4 stations. For both sets of
broadcasts, the horizontal baselines are comparable (approximately 2300 m and 3200 m for SPICE$\to$A2 and SPICE$\to$A4, and approximately 3700 m, in each case, for broadcasts from the deep pulsers). Since each station consists of
eight doublets of approximately co-located v and h receivers, each station provides up to eight
independent measurements
of the h-v arrival time difference, $\Delta T_{h,v}$. 

Using data for the v channels (notated by 0--7) versus the h channels (notated by 8--15), with v-h doublet
channel assignments separated by eight units
presented previously by \citet{tom2019ICRC}, we fitted the
$\Delta T_{h,v}$ distributions to Gaussian signal shapes. The results of fits that converged and also had an acceptable 
$\chi^2$ are presented in Table \ref{tab:fits}. We note the clear offset between the most-probable signal arrival time
differences for the A2 station (significant negative h-v time delays) versus the A4 station (generally negligible h-v time delays). Further context for how the polarization delays can be use for neutrino source reconstruction is given in Section 6.1. 
%
\begin{table}[htpb]
\begin{tabular}{c|c|c}
Pulser source & Receiver station/v-h pair & $\Delta T_{h,v}$ (ns) \\ \hline
SPICE  & A2 (0,8) & -14.03$\pm$2.35 \\
SPICE  & A2 (1,9) & -14.04$\pm$3.29 \\
SPICE  & A2 (2,10) & -14.51$\pm$3.49 \\
SPICE  & A2 (3,11) & -14.64$\pm$2.59 \\
SPICE  & A2 (4,12) & -15.09$\pm$2.29 \\
SPICE  & A2 (5,13) & -13.62$\pm$2.98 \\
SPICE  & A2 (6,14) & -13.76$\pm$2.33 \\
SPICE  & A2 (7,15) & -13.15$\pm$2.95 \\ \hline
SPICE  & A4 (1,9) & -2.06$\pm$5.66 \\
SPICE  & A4 (2,10 & 8.97$\pm$10.95 \\
SPICE  & A4 (4,11) & 7.03$\pm$10.57 \\ \hline
Deep Pulser  & A2 (0,8) & -24.49$\pm$1.40 \\
Deep Pulser  & A2 (1,9) & -27.12$\pm$2.78 \\
Deep Pulser  & A2 (2,10) & -24.57$\pm$2.27 \\
Deep Pulser  & A2 (3,11) & -27.23$\pm$2.41 \\
Deep Pulser  & A2 (4,12) & -24.71$\pm$1.35 \\
Deep Pulser  & A2 (6,13) & -24.70$\pm$2.28 \\ \hline
Deep Pulser  & A4 (0,8) & 1.55$\pm$1.30 \\
Deep Pulser  & A4 (1,9) & 1.44$\pm$1.19 \\
Deep Pulser  & A4 (2,10) & 1.60$\pm$1.56 \\
Deep Pulser  & A4 (4,12) & 2.36$\pm$1.04 \\
Deep Pulser  & A4 (5,13) & -3.23$\pm$1.74 \\ \hline
\end{tabular}
\caption{Results of Gaussian fits to $\Delta T_{h,v}$ distributions for v-h doublets, in units of nanoseconds.
Due to their very similar trajectories, broadcasts from the deep pulser sources,  IC1S DP and IC22S DP, are grouped together. $\Delta T_{h,v}<0$ corresponds to h polarization signal arrival at the receiver station before the v polarization signal arrival.
}
\label{tab:fits}
\end{table}
\section{Modelling ice birefringence}
\subsection{Effective medium model}
At radio frequencies the bulk dielectric tensor, principal refractive indices, and birefringence of polar ice can be modeled by combining ice fabric measurements (the $c$-axis orientation distribution) with information about the birefringence of individual ice crystals \citep{Fujita2006,Matsuoka2009}. This effective medium approach was developed to interpret polarimetric radar sounding measurements \citep{Fujita2006,Matsuoka2012,Brisbourne2019,Jordan2019} and assumes that the dimensions of the ice crystal grains ($\sim$ mm) are much less than radio wavelength in ice ($\sim$ dm-m for the frequency-range relevant to ARA).

Individual ice crystals have hexagonal structure and are uniaxially birefringent with the optic axis aligned with the crystallographic axis ($c$-axis) \citep{Hargreaves1978}. The magnitude of the crystal birefringence (in terms of the permittivity) is given by ${\Delta\epsilon}'={(\epsilon_{\parallel c}-\epsilon_{\bot c})}$ where $\epsilon_{\parallel c}$ and $\epsilon_{\bot c}$ are the principal permittivities parallel and perpendicular to the $c$-axis with $\epsilon_{\parallel c}>\epsilon_{\bot c}$  \citep{Fujita2000}. (The notation ${\Delta\epsilon}'$ is used as the radar sounding literature normally uses ${\Delta\epsilon}$ for the bulk birefringence.) At radio frequencies and as ice temperature increases from -60-0$^{\circ}$ C, ${\Delta\epsilon}'$ increases by $\sim$ 5$\%$ from $\sim$ 0.0325-0.0345 \citep{Fujita2000,Matsuoka1996}.

Ice fabric measurements consist of thin ice core sections that measure the ice crystal orientation distribution in terms of a second order orientation tensor \citep{Montagnat2014,Woodcock1977}. The orientation tensor represents the $c$-axis orientation distribution as an ellipsoid where the eigenvalues, $E_{1}$, $E_{2}$, $E_{3}$ represent the relative $c$-axis concentration along each principal coordinate direction. The eigenvalues have the property $E_{1}+E_{2}+E_{3}=1$, and following the radar polarimetry convention we assume, $E_{3}>E_{2}>E_{1}$. Using this eigenvalue framework, the bulk principal dielectric tensor is given by
\begin{equation}
\underline{\underline{\epsilon}}=
\left(\begin{array}{ccc}
 \epsilon_{\bot c}+E_{1}\Delta\epsilon' & 0 & 0 \\ 0 &\epsilon_{\bot c}+E_{2}\Delta\epsilon' & 0 \\ 0 & 0 &\epsilon_{\bot c}+E_{3}\Delta\epsilon' 
\end{array} \right),
\label{COFtensor}
\end{equation}
\citep{Fujita2006}. For the general case, ($E_{1}\neq E_{2}\neq E_{3}$), polar ice therefore behaves as a biaxial medium (three different principal permittivities). The principal refractive indices, which correspond to the axes of biaxial indicatrix ellipsoid are given by
\begin{eqnarray}
n_{1}=\sqrt{\epsilon_{\bot c}+E_{1}\Delta\epsilon'} \\
n_{2}=\sqrt{\epsilon_{\bot c}+E_{2}\Delta\epsilon'}\\
n_{3}=\sqrt{\epsilon_{\bot c}+E_{3}\Delta\epsilon'}.
\label{RIs}
\end{eqnarray}
\cite{Matsuoka2009} further discusses the biaxial indicatrix representation, and how it relates to different propagation directions in radar sounding.

The ice fabric depends on the stress regimes present an ice sheet. Due to ice viscosity being an order of magnitude higher parallel to the $c$-axis than perpendicular, aggregates of ice crystals tend to align toward the compression axis and away from the extension axis \citep{Alley1988}. End-member classes used to describe ice fabrics are: `random' ($E_{1}\approx E_{2} \approx E_{3} \approx \frac{1}{3}$ and associated with the near-surface),`single-pole' ($E_{1} \approx E_{2} \approx 0, E_{3} \approx 1$ and associated with deeper ice undergoing vertical compression), and `vertical girdle' ($E_{1} \approx 0, E_{2} \approx E_{3}\approx \frac{1}{2}$ and associated with lateral tension).
\subsection{Principal refractive index profiles from the SPICE ice core}
\label{S2}
\begin{figure}
\begin{center}
\includegraphics[width=8 cm]{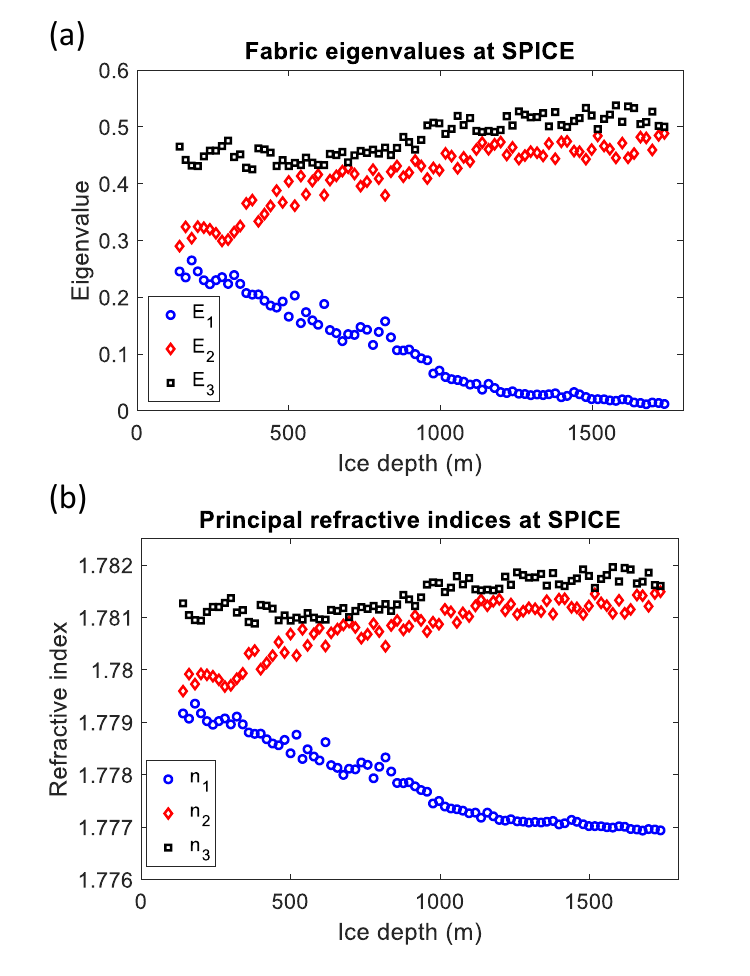}
\caption{(a) Fabric eigenvalues from the SPICE ice core \citep{Voigt2017}. (b) Eigenvalues translated into principal refractive indices.}
\label{SPICE}
\end{center}
\end{figure}

Figure \ref{SPICE}a shows fabric eigenvalue profiles from the SPICE ice core at a $\sim$ 20 m vertical resolution \citep{Voigt2017}. The plot indicates development from a relatively random fabric in shallower ice ($E_3\approx 0.45, E_2\approx 0.30, E_1\approx$ 0.25 at $z$ = 150 m) to a vertical girdle fabric in deeper ice ($E_{3}$ $\approx$ 0.50, $E_2\approx$ 0.49, $E_1\approx$ 0.01 at $z$ = 1750 m).  Profiles for the principal refractive indices, calculated as above with $\epsilon_{\perp}$=3.157 and $\Delta\epsilon'$ = 0.034  (corresponding to a polarization-averaged refractive index $\bar{n}$ = 1.78) are shown in Fig. \ref{SPICE}b. 

Due to the dominance of vertical compression, the 3 direction (greatest $c$-axis concentration) is generally close to being aligned with the vertical \cite{Matsuoka2009}, and is typically approximated as vertical in polarimetric radar sounding studies \citep{Fujita2006,Brisbourne2019,Jordan2019}. Since the azimuthal orientation of the ice core fabric sections are not recorded during drilling, the 1 and 2 directions are not initially known. For an ice flow model where there is a lateral component of tension present, the 2 direction (greatest $c$-axis concentration in horizontal plane) is expected to be approximately perpendicular to the horizontal flow direction, with the 1 direction parallel to flow \citep{Fujita2006,Wang2002}. Using polarimetric radar sounding, this predicted behavior has been verified at ice divides such as the NEEM ice core region in northern Greenland \citep{Jordan2019,Dall2010}.
\section{Modelling oblique radio wave propagation}
In the radio propagation model, we assume that the ice sheet can be modelled as a stratified anisotropic medium. Each layer of the ice sheet has thickness $\delta z_{i}$ which is determined from the vertical spacing of the SPICE ice core eigenvalue data ($\sim$ 20 m). The dielectric properties of each layer are defined by the dielectric tensor, Eqn. (\ref{COFtensor}), corresponding to the principal refractive indices, Eqn (\ref{RIs}). Lacking information about the azimuthal fabric orientation, we assume the conventional fabric orientation described in Section 3.2, with the 3 direction vertical, the 2 direction perpendicular to flow and the 1 direction parallel to flow.

\begin{figure}
\begin{center}
\includegraphics[width=9 cm]{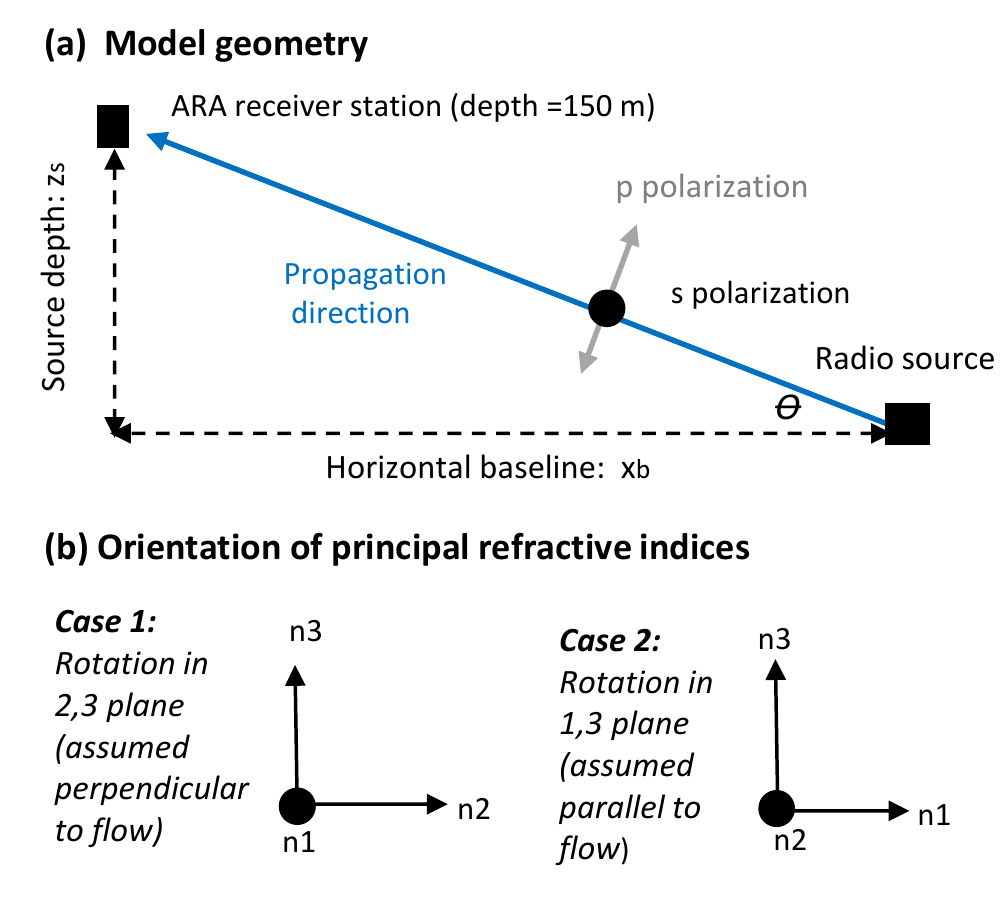}
\caption{(a) Geometry for oblique propagation model. (b) Orientation of principal refractive indices in relation to model geometry for the two cases considered. The model assumes that the $E_{3}$ eigenvector is vertical, the $E_{2}$ eigenvector is perpendicular to flow and the $E_{1}$ eigenvector is parallel to flow. }
\label{schematic}
\end{center}
\end{figure}

The radio propagation model is formulated for s- and p-polarizations (E-field perpendicular and parallel to the rotation/incidence plane, respectively); a schematic of the model geometry is shown in Fig. \ref{schematic}a. The model considers two bounding rotation scenarios: (i) rotation in the 2,3 plane (assumed to be in the plane perpendicular to flow), (ii) rotation in the 1,3 plane (assumed to be to be in the plane parallel to flow). The principal refractive indices in relation to model geometry are shown in Fig. \ref{schematic}b. For these restricted rotation scenarios, the s- and p-polarizations propagate along independent paths within the ice sheet and a double refraction/ray propagation model can be used. 

This model approach is analogous to modelling oblique propagation birefringent optical reflectors \citep{Orfanidis2016,weber2000giant}. Computationally, the model is set-up with the source depth and the horizontal baseline fixed with $\sin(\theta_{p,i})$ and $\sin(\theta_{s,i})$ as degrees of freedom to be solved for subject to Snell's law being satisfied in each layer. The model is equivalent to Fermat's least time principle being satisfied separately by each polarization mode.  

For rotation in the 2, 3 plane the s- and p-polarization refractive indices of the $i^{th}$ layer are given by
\begin{eqnarray}
n_{s,i}&=&n_{1,i}, \label{S} \\
n_{p,i}&=&\frac{n_{3,i}n_{2,i}}{\sqrt{{n_{2,i}^{2}\sin^{2}(\theta_{p,i})+n_{3,i}^{2}\cos^{2}(\theta_{p,i})}}},\label{P} 
\end{eqnarray}
where the first subscript indicates the principal refractive index component with $\theta_{p,i}$ the p-polarization propagation angle in the $i^{th}$ layer \citep{Orfanidis2016,Matsuoka2009}. For rotation in the 1, 3 plane the 1 and 2 subscripts are interchanged between Eqn (\ref{P}) and Eqn (\ref{S}). The propagation angles in each layer are derived from separate applications of Snell's law:
\begin{eqnarray}
n_{p,i}\sin(\theta_{p,i})&=&n_{p,0}\sin(\theta_{p,0}) \label{snell1} \\
n_{s,i}\sin(\theta_{s,i})&=&n_{s,0}\sin(\theta_{s,0})  \label{snell2} 
\end{eqnarray}
where $\theta_{s,i}$ is the s-polarization propagation angle in the $i^{th}$ layer and the subscript $0$ indicates the source layer. \citet{Orfanidis2016} provides analytical expressions for $\theta_{p,i}$ in terms of the principal refractive indices that were used in the model code. The deviation between $\theta_{s,i}$ and $\theta_{p,i}$ from a straight-line trajectory increases with the angle of incidence, with the highest angular offset in our simulation domain $\sim$ 0.2$^\circ$.

The radio propagation model enables calculation of the s-p signal arrival time delay for rotation in the planes of the principal axes, which serve as bounding cases for the observed h-v time delays at the receiver stations. In each layer of the ice sheet the radio path lengths are given by
\begin{eqnarray}
\delta r_{p,i}=\frac{\delta z_{i}}{\cos(\theta_{p,i})}\\
\delta r_{s,i}=\frac{\delta z_{i}}{\cos(\theta_{s,i})},
\end{eqnarray}
which corresponds to layer time increments
\begin{eqnarray}
\delta t_{p,i}=\delta r_{p,i}n_{p,i}/c\\
\delta t_{s,i}=\delta r_{s,i}n_{s,i}/c,
\end{eqnarray}
and the total s-p time delay is then given by
\begin{equation}
\Delta T_{s,p}=\sum_{i}\delta t_{s,i}-\sum_{i}\delta t_{p,i}.
\end{equation}
When $\Delta T_{s,p}<$ 0 the s-polarization arrives at the detector before the p-polarization. 
\section{Results}
\subsection{Model bounds on ice birefringence and polarization time delays}
\begin{figure*}[t]
\begin{center}
\includegraphics[width= 17.5cm]{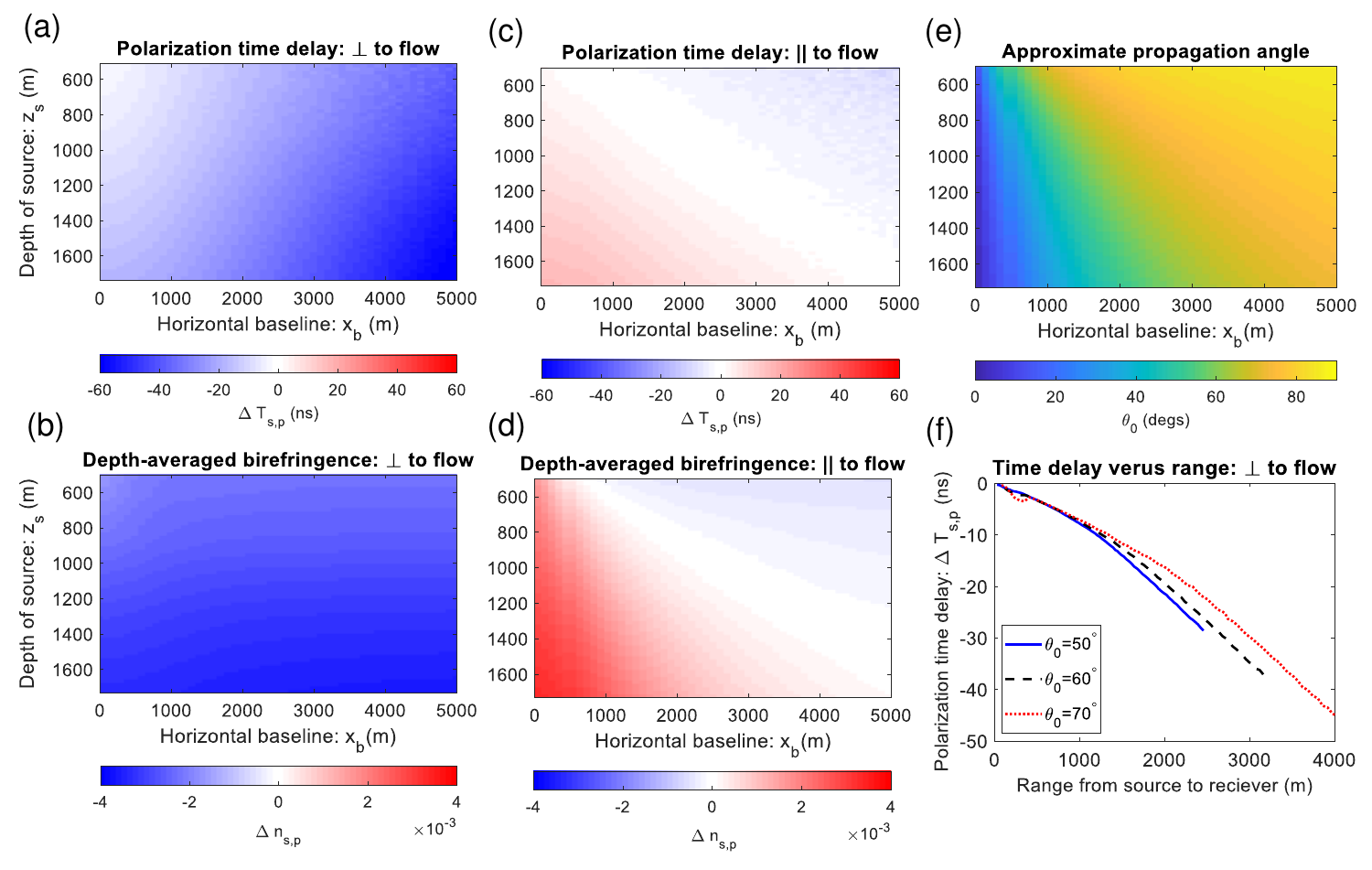}
\caption{Model bounds on ice birefringence and polarization time delays. (a) Polarization time delay for flow-parallel trajectories. (b) Depth-averaged birefringence for flow-parallel trajectories. (c) Polarization time delay for flow-perpendicular trajectories. (d) Depth-averaged birefringence for flow-perpendicular trajectories. (e) Polar propagation angle for straight-line trajectory. (f) Polarization time versus range for flow-perpendicular trajectories at fixed propagation angles. The model assumes that the $E_{3}$ eigenvector is vertical, the $E_{2}$ eigenvector is perpendicular to ice flow and the $E_{1}$ eigenvector is parallel to ice flow. The curves are different lengths in panel (f) as the range for a given depth increases with propagation angle.}
\label{Main results}
\end{center}
\end{figure*}
Figure \ref{Main results}a) and b) show modelled s-p time delays perpendicular and parallel to ice flow for a range of source depths and horizontal baselines. For the flow-perpendicular case, $\Delta T_{s,p}$ is always negative and increases with both ice depth and horizontal baseline. For the flow-parallel case, $\Delta T_{s,p}$ switches sign from positive to negative at incidence angle $\sim$ 65-70$^\circ$. $|\Delta T_{s,p}|$ is considerably more significant perpendicular than parallel to flow.

There is a simple explanation for the relationships in Fig. \ref{Main results}a) and b) in terms of the s-p birefringence: $\Delta n_{s,p}=n_{s}-n_{p}$. (In the flow-perpendicular case $n_{s}$ is given by $n_{1}$ and $n_{p}$ is given by a mixing of $n_{2}$ and $n_{3}$, whereas for the flow-parallel case $n{s}$ is given by $n_{2}$ and $n_{p}$ is given by a mixing of $n_{1}$ and $n_{3}$). Depth-averaged plots for $\Delta n_{s,p}$ are shown in Fig. \ref{Main results}c) and d), with the incidence angle for a straight-line trajectory shown in Fig. \ref{Main results}e. In the flow-perpendicular case, since $n_{1}<n_{2}<n_{3}$, it follows that $\Delta n_{s,p}<0$ and $\Delta T_{s,p}<0$ hold for all angles of incidence. In the flow-parallel case, $\Delta n_{s,p}$ switches sign from positive to negative at the same incidence angle where $\Delta T_{s,p}$ switches sign. Conceptually, this behavior occurs due to the p-polarization refractive index becoming dominated by $n_{3}$ refractive index component at oblique angles and $n_{1}$ for shallow angles. At normal incidence ($x_{b}$ = 0), $\Delta n_{s,p}$ for the flow-perpendicular case equals $-\Delta n_{s,p}$ for the flow-parallel case.

Previous analysis of ice birefringence at the South Pole is discussed in terms of fractional `birefringent asymmetry' \citep{allison2019measurement, kravchenko2011radio} which be estimated via the ratio $|n_{s}-n_{p}|/\bar{n}$ where $\bar{n}=1.78$ is the mean refractive index. For the flow-perpendicular case $|n_{s}-n_{p}|$ ranges from 0.02-0.035, which corresponds to $|n_{s}-n_{p}|/\bar{n}$ $\sim$ 0.11-0.20 $\%$. 

In the ARA experiment the elevation and azimuth angles of a trajectory are well-constrained (see Sect. 6.1 for more details), with the range (i.e. the distance along a trajectory for specified elevation and azimuth) the major unknown. Figure \ref{Main results}f) shows the relationship between the modeled polarization time delay and range for the flow-perpendicular scenario (fixed azimuth) for three fixed propagation angles (fixed elevation).
\subsection{Model-data comparison}
\begin{figure}[t]
\begin{center}
\includegraphics[width=8.6 cm]{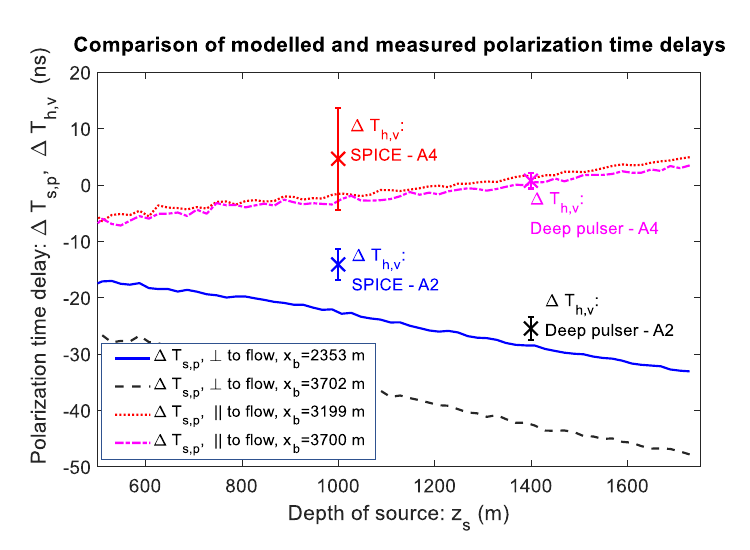}
\caption{Model-data comparison between s-p and h-v polarization time delays. The SPICE-A2 and deep pulser-A2 trajectories are approximately perpendicular to ice flow with the SPICE-A4 and deep pulser-A4 trajectories approximately parallel.}
\label{ModelData}
\end{center}
\end{figure}

The modelled s-p time delays for the flow-perpendicular and flow-parallel cases are compared with measured h-v time delays from the A2 and A4 stations (mean and rms deviation averaged over all channels in Table 1) respectively in Fig. \ref{ModelData}. This comparison assumes equivalent horizontal baselines to the source-receiver measurements and the model curves in Fig. \ref{ModelData} correspond to vertical slices in Fig. \ref{Main results}a) and c). For the flow-perpendicular case, the measured time delays are $\Delta T_{h,v}$=-14.1 $\pm$ 2.8 ns (SPICE-A2 trajectory) and $\Delta T_{h,v}$=-25.2 $\pm$ 2.0 ns (deep pulser-A2 trajectory) with modeled time delays $\Delta T_{s,p}$=-22.5 ns and -42.5 ns respectively. For flow-parallel case, the measured time delays are $\Delta T_{h,v}$=4.6 $\pm$ 9 ns (SPICE-A4 trajectory) and $\Delta T_{h,v}$=0.7 $\pm$ 1.4 ns (deep pulser-A4 trajectory) with modeled time delays $\Delta T_{s,p}$= -1.6 ns and $\Delta T_{s,p}$ =0.2 ns, respectively. 

The magnitude of the modeled polarization time delays are 60$\%$ (SPICE) and 69$\%$ (deep pulser) greater than measured for the flow-perpendicular/A2 case. The overestimate of the modeled time delay is unsurprising as the modelled flow-perpendicular scenario serves as maximum bound upon $|T_{s,p}|$. (This bound follows from the inequality $n_{3}>n_{2}>n_{1}$, and the assumption of a fixed fabric orientation with respect to ice depth). As described in Section 2, the A2 trajectories are only approximately perpendicular to flow, which would result in a reduction in the birefringence from the bounding case.

It also is to be noted that the modeled p-polarization is not strictly equivalent to the measured v-polarization.  In general, the p polarization consists of horizontal and vertical components, becoming solely vertical at glancing incidence (90$^\circ$). However, the trajectories in the model-data comparison, Fig. \ref{ModelData}, are highly oblique with the smallest angle of incidence $\sim$ 70$^\circ$.
\section{Discussion}
\subsection{Implications for range reconstruction of neutrino interactions}
The goal of experiments such as ARA is to, first, observe ultra-high energy neutrinos ($E_\nu>10^{17}$ eV) originating from outside the solar system, and, second to accumulate a sample of neutrinos with sufficient statistics to measure both a neutrino energy spectrum and also source arrival directions in the sky. In practice, as neutrinos are observed via the radio Cherenkov emissions emanating from the in-ice collision point, inferring the original neutrino direction (coincident with the central axis of the signal Cherenkov cone) requires measurement of the electric field signal polarization, projected onto v- and h-oriented receiver antennas. 

Although this study does not inform such estimates of source direction, it does, nevertheless, have
direct implications for neutrino energy reconstruction, which requires estimating the range, $R$, to an in-ice collision point. Knowledge of the range
can then be translated into a neutrino energy estimate, after correcting for $1/R$ diminution and in-ice signal absorption. Current interferometric
interaction-point reconstruction methods provide excellent angular resolution, of order 0.5 degrees in both elevation and azimuth, but
are generally unreliable for range estimates, which conventionally require measurement of the wavefront curvature across receivers separated
by ${\cal O}$(10 m), and viewing a source point over a distance of ${\cal O}$(1000 m).

The results of this study offer considerable promise for range determination, assuming the ray inclination relative to the local fabric has first been determined from interferometry. Specifically, Fig. \ref{Main results}f demonstrates that there is a near-linear negative relationship between the range and the polarization time delay for the flow-perpendicular case. As this case corresponds to maximal birefringence, it places a bound upon the minimum range that can be estimated from a given measured polarization time delay.
\subsection{Future development of the radio propagation model}
The oblique radio propagation model in this study represents simple bounding cases for polarization time delays at ARA. A key limitation is that the model is only valid for rotation planes which contain the principal axes, which results in the s- and p-polarizations propagating as independent modes through the ice sheet. A more general propagation model (formulated for a general propagation direction relative to the principal axis system) would result in wave-splitting and a coupling of the polarization modes. This mode-coupling behavior could potentially be modelled by adapting a Jones matrix model for nadir radio propagation \citep{Fujita2006} for oblique radio propagation.

The other key limitations of the model are that we assume that the fabric eigenvectors are precisely aligned perpendicular and parallel to flow and are unchanging with ice depth. Whilst the model-data comparison in Section 5.2 supports this assumption, there is likely to be at least some deviation from this idealized behavior. Joint analysis of data from all 5 ARA receiver stations, alongside the development of a more general radio propagation model, would enable better constraints to be placed upon the azimuthal fabric orientation. Polarimetric radar-sounding measurements from the ice surface could also be used as an additional constraint.
\subsection{Implications for understanding of South Pole flow history}
As discussed above, our model-data comparison supports the modelling ansatz that the $E_{2}$ eigenvector (greatest horizontal $c$-axis concentration) is perpendicular to the ice flow direction and unchanging with ice depth. This is the conventional fabric orientation that develops when a lateral component of tension is present \citep{Fujita2006,Wang2002}. By contrast, regions of the ice-sheets where pronounced flow-reorganization occurs, such as ice rises in the Weddell sea sector \citep{Brisbourne2019}, exhibit azimuthal rotation in the fabric with ice depth. 

\citet{Beem2017} inferred from radar layering and thermodynamic modelling that the South Pole region is likely have undergone recent ice flow re-organization. However, our model-data comparison is broadly consistent with there not being major fabric signatures of ice-flow re-organization (i.e. variable fabric orientation with ice depth). On the other hand, a more detailed investigation of azimuthal fabric properties and variation with ice depth could reveal more subtle fabric signatures related to ice-flow history.
\section{Summary and Conclusions}
In this study we used ice fabric data from the SPICE ice core to model radio-frequency ice birefringence and its effect upon oblique radio wave propagation relevant to in-ice neutrino detection at the South Pole. The model framework enabled us to consider bounding cases for polarization time delays across the array (perpendicular and parallel to ice flow assuming alignment with the eigenvectors of the fabric orientation tensor), with a view to placing constraints upon range reconstruction for in-ice neutrino interactions. We then compared the modelled time delays with radio-test sources measurements from the Askaryan Radio Array (ARA) neutrino experiment.

The data-model comparison demonstrated that significant polarization time delays occur for trajectories perpendicular, but not parallel, to the ice flow direction. This result can be understood from the polarization-dependent refractive indices that are modelled from the ice core fabric eigenvalue data. The model demonstrated that, for flow-perpendicular trajectories, there is an approximately linear relationship between the polarization time delay and the trajectory range. Hence, when combined with information on ray inclination, this study raises the possibility that ice birefringence can used to constrain the range to a neutrino interaction.
\section*{Acknowledgments}
TMJ would like to acknowledge support from EU Horizons 2020 grant 747336-BRISRES-H2020-MSCA-IF-2016.
DZB and AN acknowledge support from the MEPhI Academic Excellence Project (Contract No. 02.a03.21.0005) and the Megagrant 2013 program of Russia, via agreement 14.12.31.0006 from 24.06.2013.
\bibliography{Fabricbib5}   
\bibliographystyle{igs}  
\end{document}